\documentclass[sigplan]{acmart}
\usepackage{hyperref}
\usepackage[utf8]{inputenc}

\usepackage{amsmath,amssymb,amsfonts}
\usepackage{subcaption}
\usepackage{graphicx}
\usepackage{csquotes}

\usepackage{mathtools}

\DeclarePairedDelimiter\floor{\lfloor}{\rfloor}

\setcopyright{none}
\usepackage{soul}
\usepackage{tikz}
\usetikzlibrary{calc}
\renewcommand\footnotetextcopyrightpermission[1]{} % removes footnote with conference information in first column
\begin{document}

%
% The "title" command has an optional parameter, allowing the author to define a "short title" to be used in page headers.
\title{CDAG: A Serialized blockDAG for Permissioned Blockchain}

\author{Himanshu Gupta}
\affiliation{%
  \institution{Department of CSE \\ Indian Institute of Technology Madras}
%   \streetaddress{dfg; sajfd}
%   \city{Chennai}
%   \country{India}
}
\email{himanshg@cse.iitm.ac.in}

\author{Dharanipragada Janakiram}
\affiliation{%
  \institution{Department of CSE \\ Indian Institute of Technology Madras}
%   \streetaddress{dfg; sajfd}
%   \city{Chennai}
%   \country{India}
}
\email{djram@iitm.ac.in}

%
% By default, the full list of authors will be used in the page headers. Often, this list is too long, and will overlap
% other information printed in the page headers. This command allows the author to define a more concise list
% of authors' names for this purpose.

%
% The abstract is a short summary of the work to be presented in the article.

\begin{abstract}
        Blockchain is maintained as a global log between a network of nodes and uses cryptographic distributed protocols to synchronize the updates. As adopted by Bitcoin and Ethereum these update operations to the ledger are serialized, and executed in batches. To safeguard the system against the generation of conflicting sets of updates and maintain the consistency of the ledger, the frequency of the updates is controlled, which severely affects the performance of the system.
        
        This paper presents Converging Directed Acyclic Graph (CDAG), as a substitute for the chain and DAG structures used in other blockchain protocols. CDAG allows multiple parallel updates to the ledger and converges them at the next step providing finality to the blocks. It partitions the updates into non-intersecting buckets of transactions to prevent the generation of conflicting blocks and divide the time into slots to provide enough time for them to propagate in the network. Multiple simultaneous updates improve the throughput of CDAG, and the converging step helps to finalize them faster, even in the presence of conflicts. Moreover, CDAG provides a total order among the blocks of the ledger to support smart contracts, unlike some of the other blockDAG protocols.

        We evaluate the performance of CDAG on Google Cloud Platform using Google Kubernetes Engine, simulating a real-time network. Experimental results show that CDAG achieves a throughput of more than 2000 transactions per second and confirms them well in under 2 minutes. Also, the protocol scales well in comparison to other permissioned protocols, and the capacity of the network only limits the performance.

\end{abstract}

% <concept_id>10010520.10010521.10010537.10010540</concept_id>
% <concept_desc>Computer systems organization~Peer-to-peer architectures</concept_desc>
% <concept_significance>300</concept_significance>
% <concept_id>10003752.10003777.10003789</concept_id>
% <concept_desc>Theory of computation~Cryptographic protocols</concept_desc>
% <concept_significance>100</concept_significance>

% \ccsdesc[300]{Computer systems organization~Peer-to-peer architectures}
% \ccsdesc[100]{Theory of computation~Cryptographic protocols}
%
% Keywords. The author(s) should pick words that accurately describe the work being
% presented. Separate the keywords with commas.
% \vspace*{-0.1in}

% This command processes the author and affiliation and title information and builds
% the first part of the formatted document.
\maketitle

\section{Introduction}
    Bitcoin, proposed in 2008 by a pseudonym named \textit{Satoshi Nakamoto} \cite{nakamoto2008bitcoin}, introduced the first ever decentralized framework to manage digital currency over a peer-to-peer network. It stores transactions in a publicly available ledger called the \textit{blockchain}. Each block in the blockchain contains a batch of transactions referring back to the latest known block, constructing a chain of blocks. The consistency of the ledger comes from the serial extension, as the newly added block is legitimate only if it is consistent with the chain. After Bitcoin other cryptocurrencies like Ethereum \cite{wood2014ethereum}, Litecoin \cite{litecoin} originate extending the functionality of Bitcoin but using the same underlying blockchain structure. 
    
    To maintain the consistency of the ledger block generation rate is kept reasonably low because a higher rate can result in forks and increase the risk of a double spend. Also, creating larger blocks with more number of transactions results in higher propagation time, which in turn refrains the system to scale. \cite{decker2013information,sompolinsky2015secure} discuss in detail about the security problems that originate because of high block generation rate or increasing the block size for the blockchain.
    
    BlockDAG~\cite{blockdag} is introduced as an alternate structure to address the issue of the scalability of blockchain. It overcomes the problem of low transaction throughput by supporting very high block generation rate and arrange the proposed blocks in a Directed Acyclic Graph instead of a chain. However, the performance still degrades in the presence of conflicting blocks because of their sophisticated transaction confirmation strategies. Also, due to high block generation rates, the probability of concurrent blocks having the same transactions is high. It is indistinguishable from the case of having blocks with double-spend transactions since in both the scenarios, only one of the proposed blocks gets accepted for the final ledger and other needs to be discarded. Thus, the orphan rate for DAG can still have high values, even in the absence of any conflicting transactions in real life deployments.
    
    This paper describes \textit{Converging Directed Acyclic Graph} (CDAG), a new structure to maintain a consistent and distributed ledger over a peer-to-peer network. CDAG is a hybrid structure of blockchain and blockDAG, designed to achieve the best of both worlds. It progresses serially similar to the blockchain and can handle multiple simultaneous blocks analogous to the blockDAG. Splitting the transactions into non-intersecting buckets enable the block proposers to generate multiple non-conflicting blocks simultaneously, and dividing the time into slots with the help of a \textit{Distributed Time Barrier} grants them enough time to disseminate in the network. 
    
    Blocks in CDAG converge after each step into a single block known as the \textit{Converging Block} (C-Block) which is the point of reference for the next set of blocks. Adding C-Block at each step enables the ledger to progress as a chain and maintain a total ordering among the blocks. However, temporary forks can occur because the ledger progresses as a chain. CDAG handles it by using a variant of the heaviest chain protocol \cite{sompolinsky2015secure} to confirm transactions irreversibly in an acceptable amount of time.
    
    The rest of the paper is structured as follows. Section 2 discusses the related work. Section 3 provides a summary of the previous work extended in this paper. Section 4 describes the design of the protocol. Section 5 and 6 discuss the details of CDAG and describe the block proposal mechanism. Section 7 provides a security analysis of the protocol. Section 8 and 9 tells about the implementation and evaluation of the protocol. Finally, we discuss the limitations and indicate future directions in Section 10, and conclude our work in Section 11.
    
\section{Related Work}
        \subsection{Chains}
            The chain is a fundamental and most commonly used structure for the blockchain. Blocks of transactions extending the chain are added periodically one after the other, linked cryptographically to their previous blocks. Nodes mine these blocks for the chain to get the associated reward. 
            Bitcoin uses proof-of-work often called as ``Nakamoto consensus" to mine blocks, named after the creator of bitcoin. Participants try to solve a hard cryptographic puzzle, and the first one to get through proposes the block. Nodes require massive infrastructure to participate, and the problem is set too hard such that probabilistically only a single node can generate a block every 10 minutes. It incurs a lot of resource wastage, and throughput of the system still suffers. Moreover, due to the possibility of forks, users wait for the chain to grow at least six blocks before considering a transaction confirmed \cite{bitcoinConfirmation}, i.e., an hour of waiting. Many follow-up cryptocurrencies like \cite{ethereum,litecoin,zcash} adopted the bitcoin's proof-of-work and inherited its limitations.
           
            CDAG, on the contrary, allows a set of nodes in each round to propose blocks of transactions simultaneously from distinct buckets. Valid proposals from different proposers are added to the ledger in parallel, extending the ledger as a converging DAG, unlike a chain of blocks used in Bitcoin. It helps CDAG to gain efficiency and increase the throughput of the system.
            
            Bitcoin-NG~\cite{eyal2016bitcoin} takes a step forward to use proof-of-work to elect a leader for a time epoch, and let that leader propose blocks until the next leader gets elected. It improves the performance, but a single leader for each epoch is susceptible to an attack. CDAG also progresses in time slots, but there is no unique leader to propose blocks. Multiple proposers simultaneously try to introduce blocks from different buckets. It prevents the system against denial of services (DOS) attacks and ensures the progress of the protocol.

            Algorand \cite{gilad2017algorand} use anonymous committee selection to propose blocks and perform voting to select the one amongst them to be added to the ledger. It utilizes network bandwidth to propagate $n$ number of blocks, but finally adds only one amongst them to the ledger. It results in under-utilization of the network resources and compromises on the achievable throughput. CDAG try to maximize the throughput by allowing the blocks from multiple proposers to get added to the ledger concurrently. It tries to exploit the network resources and also gains on throughput.
           
        \subsection{Trees and DAGs}
            Other than chains, structures like trees and directed acyclic graph (DAG) have gained eminence lately. GHOST~\cite{sompolinsky2015secure} modifies the chain selection rule of Bitcoin to select the heaviest sub-tree instead of longest chain at a fork. It selects the branch with the highest mining power concentration, providing better security guarantees than original bitcoin, with higher block generation rates and improved fairness. CDAG also grows like a tree with multiple child blocks at each step, but they converge at the next step trying to commit a maximum possible of them.  The protocol tries to add all the non-conflicting child blocks to the ledger, unlike \cite{sompolinsky2015secure} where only a single child block ends up in the final chain. It results in better performance and supports even higher block generation rates. 
        
            Spectre~\cite{sompolinsky2016spectre} uses directed acyclic graphs as the underlying structure for the ledger instead of chains for increasing Bitcoins throughput. Blocks in \cite{sompolinsky2016spectre} points to all the leaf blocks visible to the proposer and resolve conflicts based on the number of direct and derived links for the concerned blocks. It supports a very high block generation rate in comparison to the original bitcoin blockchain and also scales easily without compromising on the security. However, Spectre provides only pairwise ordering between the blocks and not total-ordering which makes it unsuitable for \textit{Smart Contracts} \cite{szabo1997idea} based blockchain deployments like Ethereum. Moreover, it is possible that a block added in the DAG may not ever get confirmed in the presence of conflicting transactions \cite{dagPerformance}. CDAG, on the other side, converges by design at each step and thus supports smart contracts. Conflicts can occur in the form of forks but are resolved irreversibly as long as a majority of nodes are working towards extending the heaviest branch.
            
            Phantom~\cite{sompolinsky2018phantom} is also a blockDAG based protocol that provides a total ordering among the blocks added to the ledger. It makes Phantom suitable for smart contracts, but the performance degrades severely in the presence of conflicting blocks. Moreover, in both \cite{sompolinsky2016spectre,sompolinsky2018phantom} there is no mechanism for the proposers to generate distinct blocks. Multiple blocks proposed at small intervals possess a high probability of having conflicts, which can result in higher orphan rate (percentage of the proposed blocks not included in the ledger) in real-time systems. CDAG addresses this issue by dividing the unconfirmed transactions pool into multiple buckets. Multiple proposers randomly select buckets to propose blocks, reducing the probability of the generation of conflicting blocks and therefore decreasing the orphan rate.

\section{Background}
    CDAG is a structure to store blocks of transactions proposed for a ledger. It can accommodate multiple blocks proposed simultaneously but requires a distributed consensus protocol to select the proposers for each round. CDAG uses Colosseum to select block proposers for each of the time slot and add the blocks generated by them to the ledger.
    
    \subsection*{Colosseum}
        %cite colosseum here
        Colosseum~\cite{gupta2019colosseum} is a scalable consensus protocol to elect a random subset of nodes from a peer-to-peer network. It is designed for permissioned blockchain networks where the identity of participants are known. Colosseum follows a curative approach towards consensus rather than a preventive approach followed by all the voting based algorithms like \cite{castro1999practical}. It assumes that adversaries are not always present in a network, but efficiently detects them and alerts the network; whereas the traditional Byzantine Fault Tolerant consensus algorithms work on the assumption that traitors are always present in the network. Such a design assumption brings down the message complexity of Colosseum and allows the protocol to scale. 
        
        Colosseum arranges the participants in a structured ring network forming a Distributed Hash Table (DHT) \cite{dht}. These participants engage in \textit{knockout} tournaments to win and get elected as block proposers. A tournament in Colosseum consists of $\log_2 N$ ($N$ is the number of participants) asynchronous rounds. Nodes play matches in pair of two for each round and accept the winners of $\alpha$ ($\alpha < \log_2 N$) number of rounds as the proposers for that tournament. The elected proposers can then propose a block for the blockchain and propagate them in the network along with their \textit{Proof-of-Win} (PoWin). PoWin is a cryptographically secured, tamper-proof certificate designed to get communicated and stored as the outcome of matches in Colosseum. The receivers validate the PoWin and the proposed block and push it into the ledger.
        
        Members of Colosseum compete in a fair and randomized two-player game to select the winner and rely upon arbitrary members on the ring to validate and verify the result. Each match in the tournament consists of four stages.
        First, a node finds an eligible competitor for the match by sending TCP requests \cite{postel1981transmission} to other members. Eligible players respond positively providing the PoWin of their previous round while others acknowledge negatively.
        Second, players need to create their \textit{Game Proposal} for the two-player game. Cryptographic techniques are used to ensure that the nodes generate random but verifiable game proposals. It helps to have a fair winner as none of the participants can falsify the process.
        Third, competitors send their proposals to a randomly selected \textit{validator} on the DHT, which is unambiguously identifiable for a particular match. It validates the proposals of both the players and constructs the resulting certificate (i.e., PoWin) by comparing the proposals.
        Fourth, the result is sent back to the players and their \textit{keepers} by the validator. Multiple keepers are assigned to both the participants of a match to store their state on the network. These keepers follow a weak consistency approach where the validator sends the results to some random keepers and the keepers then gossip it to other keeper nodes. Moreover, nodes on receiving the result also forward it to the keepers to maintain the consistency among all the keeper replicas. 
       
        Finally, winners move forward to play the next round or propose a block if they meet the qualifying condition (i.e., have won $\alpha$ number of rounds), whereas the losers wait for the start of the next tournament. Simultaneously, keepers verify the results received by them with the keepers of the previous round. It is required to make sure that only the winners of a round proceed to the next round and vote against the players playing unauthorized matches. A match on receiving more than two-third of negative votes from its keepers is considered as a \textit{foul}. These fouls impact the priority of the blocks proposed by nodes and lowers their chances of being in the final ledger.
        
\section{Design}
        
        Converging Directed Acyclic Graph (CDAG) is proposed as an alternate data-structure to store data of a distributed ledger across a peer-to-peer network. It is similar to the traditional blockchain as it can be visualized as a chain and inspired from blockDAG to handle multiple simultaneous blocks. CDAG introduces Converging Block (C-Block) as a new construct to the structure of the ledger. C-Block is a converging point for the blocks in CDAG unlike other blockDAG protocols which do not have any single point of convergence (i.e., a block such that others are either before it or after it, making it a confirmation point for all previous blocks and last point of reference for the future blocks). Moreover, CDAG is introduced as an easy to understand and maintain structure, with more straightforward conflict resolution technique, which makes it desirable for blockchain systems.
        
        Timeline in CDAG is divided into slots enforced by a \textit{distributed time barrier} (DTB). Each tournament of Colosseum directly maps to one of the slots created by DTB.
        Multiple block proposers selected in one slot extend CDAG by proposing blocks from random buckets of transactions. They make the blocks consistent with the previous heaviest branch and disseminate in the network. The \textit{barrier} between the two time slots provides adequate time for the proposed  blocks to scatter in the network to increase their chances of converging at the next step; whereas bucketing of the transactions prevents them from generating intersecting blocks (blocks with the same transactions) because all of them have the same set of unconfirmed transactions.
        
        % conflict resolution is simple therefore performance does not degrade a lot

        % \subsection{Boot Straping}
        
        % \noindent\textbf{Joining the Network.} The system consists of trusted oracles to provide bootstrap information. Nodes can randomly poll an oracle to join the structure. Oracle will validate the node in accordance with the notion of membership, render the public-key certificate~\cite{lyons2000federal} (certificate including a unique id and public-key of user digitally signed by the oracle) and other required information to join the ring network. Such a trusted public-key infrastructure is required to verify message origin and restrict malicious nodes to masquerade others.\\

        \subsection{Distributed Time Barrier} 
        \begin{figure}
            \centering
            \includegraphics[width=\columnwidth,keepaspectratio]{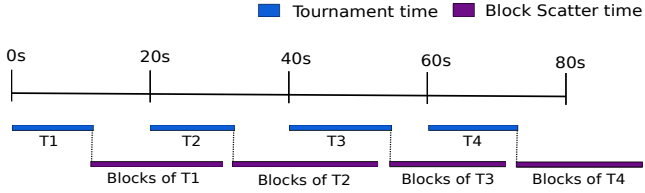}
            \caption{Timeline in CDAG with $\tau = 20s$}
            \label{fig:db}
        \end{figure}
        As the name suggests, \textit{Distributed Time Barrier} (DTB) is a time-based barrier to force the nodes of a network to halt for a particular time. Having a barrier allows CDAG to generate multiple simultaneous blocks since they will get the required time to scatter, and also restrict the nodes from moving ahead and generate more blocks until allowed. It plays a significant role in the convergence of CDAG at each stage, because, better the blocks propagate in the network more is their chance to get added to the ledger.
        
        DTB maps each time slot exclusively to a tournament of Colosseum. Nodes need to wait for $\tau$ seconds after the start of a tournament to play the next tournament. Figure~\ref{fig:db} shows the division of time into slots of 20 seconds and each slot initiating a new tournament. Nodes participate in these tournaments to qualify and propose blocks for the ledger. The proposed blocks can propagate until the proposers from the next tournament get selected, as shown in figure~\ref{fig:db}. It gives them approximately $\tau$ amount of time to scatter in the network irrespective of the time of a tournament but is valid only if the average completion time for the tournaments is less than $\tau$ and multiple consecutive tournaments do not overlap.
        
        CDAG uses Proof of Elapsed Time (PoET)~\cite{poet} to enforce waiting on the nodes. PoET with the help of Intel Software Guard Extension~\cite{sgx} provides a framework to run trusted code on an untrusted host, which others can verify. Nodes need to provide the PoET certificate of waiting $\tau$ seconds after the start of tournament $t$ to play tournament $t+1$. Each certificate contains a hash used to initialize the waiting for the next slot. These hashes are linked to each other such that nodes can start waiting for a particular tournament only after the successful generation of the PoET certificate for the previous tournament. Competitors verify these certificates while pairing to decide whether to play a match or not. It restricts the participants from moving ahead of others and thus fulfill the requirement of a time barrier across the network. The barrier makes them advance synchronously tournament-by-tournament and grants equal chance for the blocks of all the proposers of a tournament to reach a majority of the users.
        
        However, nodes in CDAG need not be completely synchronous. Time difference of a few milliseconds (500 - 1000 ms) between the participants does not affect the tournaments since $\tau$ is set sufficiently high. Still, as the process of waiting at the barrier is independent, clocks of players can drift apart further. Such nodes lacking behind need to poll the trusted oracles present in the network to reinitialize the state of their barrier. These trusted oracles also wait at the barrier along with the network and provide the bootstrap information for the trusted code precisely at the time of the start of the next slot. It also works similarly for the newly joined nodes in the system to get the bootstrap information and synchronize with others.

        \subsection{Bucketing of transactions}
        
        \begin{figure}
            \centering
            \includegraphics[width=22em,keepaspectratio]{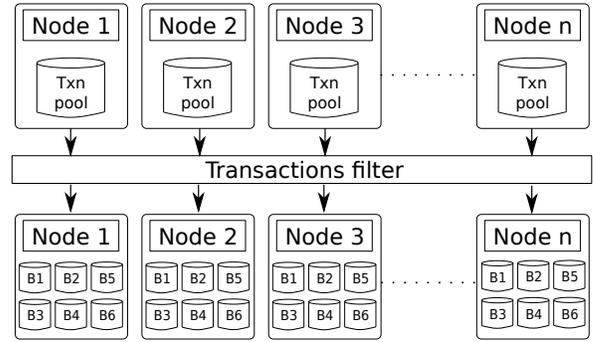}
            \caption{Bucketing of transactions}
            \label{fig:bucket}
        \end{figure}
        
        CDAG generates multiple blocks in each slot concurrently, and there is a high probability of the same transactions being included by multiple proposers in their blocks because of a single pool of unconfirmed transactions. Proposers can unknowingly produce conflicting blocks even in the absence of double-spending transactions and can increase the orphan rate (the percentage of proposed blocks not included in the final chain) of the protocol. \cite{sompolinsky2016spectre,sompolinsky2018phantom} encounters the same issue because of their high block generation rate.
        
        Transactions in CDAG are distributed among different buckets to counter this problem. The unconfirmed transactions are split across $B$ non-intersecting transaction pools by the nodes. Each transaction maps to a single bucket based on its hash, as shown in equation ~\eqref{eq:zero}. Proposers randomly select a bucket and include transactions only belonging to the particular bucket to generate a valid block. Blocks that belong to different buckets are assured of having different transactions and thus, limits the probability of the generation of conflicting blocks. The number of buckets is kept sufficiently high from the expected number of proposers to avoid any collisions in bucket selection. Also, proposers propagate the block headers in the network before sending the actual block to inform other proposers about the choice of the bucket. Block headers are comparatively smaller than the actual block and spread faster across the network, further minimizing the chances of creating conflicting blocks.
        
            \begin{equation}
                    Bucket\ Id = (Transaction\ Hash)\ mod\ B
                    \label{eq:zero}
                \end{equation}
        
        However, dividing transactions merely by their hashes do not protect the system against double-spending conflicts. Two transactions trying to spend the same money can have different hashes and may end up in different buckets. Therefore, blocks from different buckets can also conflict. 
        CDAG handle such conflicts while converging these blocks and makes sure that it does not add multiple blocks containing double spending transactions to the ledger. Thus, the protocol works on its full potential as long as the users do not double-spend and can resolve the conflicts efficiently in comparison to others by using the heaviest branch selection mechanism of CDAG.
        
\section{Converging DAG}    
    \begin{figure}
            \centering
            \includegraphics[width=\columnwidth,keepaspectratio]{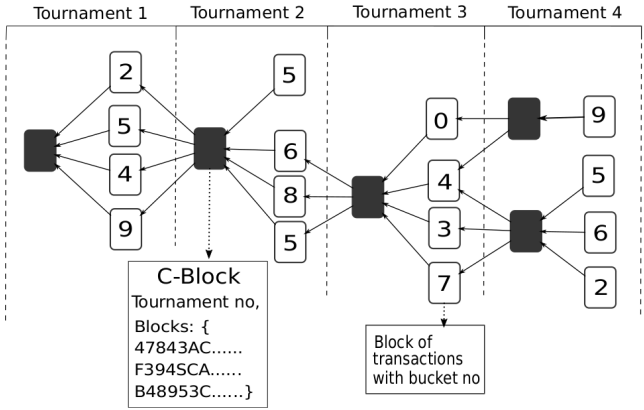}
            \caption{Structure of CDAG}
            \label{fig:cblock}
        \end{figure}     
        
    CDAG is a blockDAG based distributed ledger which converges at each step to provide finality to the blocks. It is a serialized DAG designed to support smart contracts and quickly arrive at conclusions in the presence of conflicting transactions. It tries to maximize the throughput of blockchain systems and better utilize network resources.
    
    CDAG uses Colosseum to elect multiple block proposers for each time slot. These proposers select a random bucket to generate blocks consistent with a \textit{Converging Block} (C-Block) and disseminate it into the network. CDAG constructs C-Block as a collection of the blocks proposed in a round and works as a single point of reference for the newly generated ones. It enables CDAG to progress as a chain and shares the same trust model as the traditional blockchain (blocks of transactions linked to each other progressing as a modifications resistant ledger). Progress as a chain can create temporary forks, and thus, CDAG uses the heaviest chain selection approach to have faster conclusions.
    
    % Temporary forks can occur in the chain and therefore, CDAG uses heaviest chain selection approach and introduce an algorithm to confirm transactions irreversibly in CDAG.
    
    % A block consistent with a C-Block is bound to be consistent with all the blocks included in it.
    
    \subsection{C-Block}
           
            \textit{C-Block} (Converging block) is the converging point in CDAG. It is introduced as an extra layer between two layers of blocks providing a point of convergence to the blocks of the previous slot and a single point of reference for all the blocks of the current slot, shown in Figure~\ref{fig:cblock}. As multiple blocks are recommended for the ledger in each slot, a C-Block refers to various non-conflicting blocks and contains the list of hashes of all the previous blocks it refers. A block consistent with a C-Block is bound to be consistent with all the blocks included in it.
            It allows the current block proposers to refer multiple blocks with a single link and enables different proposers to have the same view about the ledger. 
            
            Creation of C-Block is also the responsibility of the block proposers. While proposing a block, a proposer needs to make it consistent with a C-Block and includes a pointer to it. To do so, a proposer can either generate a new C-Block or use an existing C-Block created and propagated by other selected proposers of the same tournament. In case of a conflict, the heaviest chain rule is used to select the C-Block with higher probability to survive.
            
            Generating a valid C-Block is a trivial process, and obeys a specific set of rules. Proposers need to make sure that all the blocks included in a C-Block must: 1) belong to different buckets, 2) have non-conflicting transactions, 3) refer to the same previous C-Block. The first rule ensures that a C-Block does not contain conflicting blocks due to the presence of the same transactions. Second does not allow double spending transactions in a C-Block, and third provides the serial nature to CDAG. Blocks of a tournament with different previous C-Blocks are similar to a fork in the traditional blockchain. In figure~\ref{fig:cblock} the block of bucket 9 points to a different C-Block than that of buckets 5, 6, 2 at tournament 4, i.e., the previous blocks of both the C-Blocks are different, creating a fork. Thus, the structure progresses as a chain where a C-Block can be seen as a large block encapsulating all the blocks included in its list. 
        
            % Moreover, a C-Block of tournament $t$ can only contain the hashes of the blocks proposed in previous tournaments and is referred by the blocks of tournament $t$, defining its position in CDAG visible in Figure~\ref{fig:cblock}.
            
    \subsection{Total ordering for Smart Contracts}
        CDAG is designed as an alternate to other blockDAG based blockchain protocols to incorporate smart contracts. Phantom \cite{sompolinsky2018phantom} also supports smart contracts by providing ordering between the blocks but is not as intuitive as provided by CDAG. Blocks in CDAG are ordered at two levels, and together it ensures a total ordering among all the blocks added in the main branch of the ledger, similar to a bitcoin blockchain. First, the arrangement of C-Block provides a higher level order in CDAG. Since CDAG extends as a chain of C-Blocks, blocks included in the predecessor (C-Block) of a block are inherently behind the current block. Second, a lower level ordering is provided in between the blocks included in a C-Block. These blocks are ordered based on their block hash resulting in an inter-block ordering. Such blocks do not refer to each other but are non-conflicting by nature as they are present in the same C-Block. Together both the properties provide the required total ordering at block level among the blocks of CDAG.

    \subsection{Heaviest chain selection}
            
            \begin{figure}
                \centering
                \includegraphics[width=3.1in,keepaspectratio]{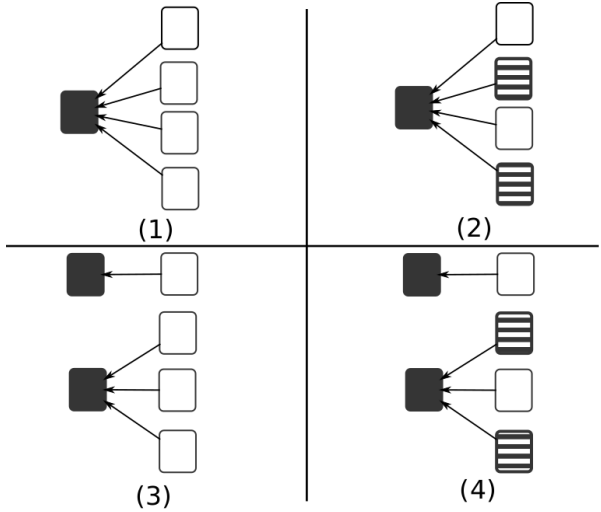}
                \caption{Possible leaf blocks formation while the generation of C-Block where Blocks with horizontal stripes conflict with each other.}
                \label{fig:conflict}
            \end{figure}
            
            Heaviest chain selection in CDAG follows a greedy approach towards extending the chain with the highest probability to be the main chain. \cite{sompolinsky2015secure} motivates the chain extension algorithm used by CDAG, with a difference of trying to extend the chain of C-Blocks instead of regular blocks. The base assumption for the algorithm is that a majority of nodes in the system are honest and try to identify and report any unauthorized activity. It relies on the negative voting done against malicious nodes in Colosseum to identify honest blocks and branches. Votes against nodes reduce the priority of their blocks and therefore reduce its chance to be included in the main chain.
            
            Each block in CDAG has an associated weight $w$ which depends on the number of successfully verified matches of its proposer. Maximum weight for a block is $\alpha$ corresponding to the $\alpha$ number of wins gained to propose it. \textit{Foul} (a match that is known to be unauthorized because of majority voting against it by the keepers of Colosseum) by nodes negatively affect the weight of these blocks. Each foul decrease the weight of the corresponding block by one unit resulting in low priority. Participants consider these votes while the generation of C-Block and decide whether to include a block or not.

            The total weight of a chain $W$ is the sum of the weights of all the blocks present in it. Nodes use it at two decision-making points during the tournaments. First, when a node needs to propose a block and has multiple C-Blocks to consider as a reference. It is similar to a fork in the blockchain, and the proposer needs to select the chain with the highest likelihood to survive. Second, when proposers need to generate a C-Block. Proposers need to maximize the weight of the chain they are proposing their block to by including the heaviest subset of non-conflicting blocks from the set of proposed blocks. In such a situation there can be multiple cases as shown in figure \ref{fig:conflict}, 1) A single chain with no conflicting blocks; 2) A single chain with conflicts; 3) A forked chain with none of the chain having conflicting proposals; 4) A forked chain with conflicting block in both the chains or one of the chain.
            The first case is simple, where the proposer includes all the blocks while constructing the C-Block. In the second case, the proposer needs to select the heavier block among the conflicting blocks to include to the chain. In the third case, the proposer selects the heaviest of the multiple forks to construct the ledger, whereas in the fourth case the proposer first sorts the conflicts in-between the forks and then select the heaviest branch amongst them. 
            Moreover, in the case when multiple conflicting blocks have the same weight, the block of the proposer with a smaller Id is considered. It ensures uniform generation of C-Block across the network, given all of them have received the proposed blocks and the negative votes by the keepers.
    
            % Participants do not consider blocks proposed by the proposers with one or more of their $\alpha$ matches known as fouls to be added to the chain, preventing random nodes from proposing blocks.
            
\section{Block Proposal}
        
        Nodes with $\alpha$ number of wins in a tournament of Colosseum are eligible to propose a block for CDAG.  First, they select the heaviest C-Block to mark it as a reference in the new block, then proposes a block referencing the selected C-Block as the previous block. It attaches its latest PoWin (of the round number $\alpha$) in the block header and propagates it into the network. 

        Each tournament of Colosseum can have a maximum of $\log(N)$ ($N$ is the number of nodes) rounds; therefore, $\alpha$ is always set as $ \alpha < \log(N)$ to have multiple block proposers for a tournament. 
        CDAG tries to commit all the non-conflicting blocks simultaneously using CDAG. It helps to utilize the bandwidth of the network better and decrease resource wastage. Blocks proposed by different nodes from distinct parts of the network propagate simultaneously, as compare to a single block getting propagated. It allows them to scatter in the entire network in almost the same amount of time as a single block without any extra cost. The distributed time barrier across tournaments also supports the design decision as it provides adequate time for the blocks to scatter in the network.

        \subsection{Multiple blocks per tournament}
            Allowing multiple block proposers to propose concurrently can escalate the generation of conflicting blocks because of the same unconfirmed transaction pool. Blocks can conflict in two ways. First, when two different blocks have an intersection (i.e., contains the same transactions). Second, when there is a double-spend across two blocks. The second one is a deliberate attempt, but the first issue occurs only because of the delays in the network. A proposer not aware of the blocks proposed by others can unknowingly generate a conflicting block. 
            
            CDAG uses bucketing of transactions to minimize the probability of the generation of intersecting blocks. However, bucketing does not resolve the conflicts that occur due to double-spend transactions. Such conflicts are resolved during the generation of C-Block by the block proposers of the next tournament. A C-Block contains the pointers to the previous blocks and is the point of reference for the newer blocks (similar to a previous block in blockchain). A valid C-Block does not contain pointers to multiple conflicting blocks. Thus, as long as a majority of the nodes extends the heaviest chain with valid C-Blocks, a double-spend cannot be successful.
            
            Generating multiple blocks simultaneously also elevate the problem of having total ordering between the blocks. Blocks introduced concurrently or at small interval are unaware of the other proposals and do not align in an order. The design of C-Block handles this problem and also provides order to the structure. Since a C-Block collects only non-conflicting blocks into it and their hashes are arranged in sorted order, multiple blocks included to CDAG in a single slot are inherently ordered. Moreover, a higher level order is provided by the arrangement of CDAG as a chain of C-Blocks resulting in a total order among the blocks.
            
        \subsection{Waiting time for Distributed barrier}
            Each participant of the network waits for $\tau$ amount of time after the start of a tournament to play the next tournament. Choice of $\tau$ severely impacts the performance of the protocol since multiple blocks are being generated simultaneously. A small value of $\tau$ does not provide enough time for the proposed blocks to reach everyone in the network. It can lead to a fork in the chain as there exist multiple combinations in which a node can commit blocks to the ledger. Proposers of the next slot will propose blocks according to their view of the network, which can lead to an increase in the orphan rate of the protocol and also affect the latency. Hence, CDAG prefers to have a large enough value of $\tau$ because with multiple blocks per tournament and not enough time to propagate forking can be harmful to the safety of the protocol. It is also a significant reason to impose waiting on the nodes explicitly with the help of distributed time barrier.
            
            The value of the waiting time ($\tau$) depends on external as well as internal factors. External factors are the parameters that do not depend on the protocol like bandwidth, the number of nodes and internal factors are the tunable parameters of the protocol, such as $\alpha$ and block size. $\tau$ is directly proportional to the number of nodes and block size. The number of nodes in the network and the size of the block both increase the time required for the blocks to scatter in the network. Larger network size increases the number of matches played in the tournament to qualify which also requires more time.
            $\tau$ is inversely proportional to the network bandwidth as higher bandwidth decreases the tournament completion and block scatter time.  However, the relation between $\tau$ and $\alpha$ is not straightforward. The time required for higher values of $\alpha$ increases because of the increase in the number of matches per tournament, and decrease because the number of qualifying nodes per tournament will drop, resulting in fewer blocks. Indirectly it depends on the external factors (i.e., bandwidth and number of nodes) to decide whether the overall effect is positive or negative.     

\section{Security Analysis}
    Securing blockchain systems is a significant concern. Malicious nodes can try to undermine the security of the system for their gain. Therefore, a blockchain protocol must be able to tackle adversaries and continue to work as long as the majority of the nodes in the system are honest. This section reviews such problems and discusses the countermeasures adopted in CDAG and Colosseum. 
    
    \subsection{Irreversible transaction confirmation} \label{txnCnfrm}
            Transaction confirmation is an essential aspect for all the blockchain systems. A transaction, once confirmed, must be irreversible to avoid a double spend. In bitcoin blockchain, confirmation of a block depends on its depth in the chain, but CDAG has a subtle difference. Multiple blocks get added to CDAG in each time slot, and the deepest chain may not be the one to which the majority of the participants are contributing. 
            Therefore, Colosseum introduces the concept of \enquote{\textit{full-confirmation}} for CDAG. A block is said to receive one full-confirmation when it has $\delta$ (maximum expected blocks or block proposers from a tournament) number of blocks ahead of it. Equation~\eqref{eq:five} shows how to compute $\delta$ for a given network where $N$ is the number of nodes and $\alpha$ is the number of wins required in Colosseum to propose a block.
                
                \begin{equation}
                    \delta = \floor*{\frac{N}{2^{\alpha}}}
                    \label{eq:five}
                \end{equation}
                
            A block can get a full-confirmation in a minimum of one and a maximum of $\delta$ number of tournaments. It is because the maximum number of blocks that can be added to the ledger in a slot is $\delta$ and the minimum is $1$. The blocks proposed in a tournament can have conflicts or may not reach everyone in the given time slot due to network delays resulting in a variable number of blocks getting added to the ledger. In such cases, a full-confirmation may span up to multiple tournaments.
            
            In CDAG, a block is considered as confirmed when it receives $x$ full-confirmations in less than $2x$ number of tournaments. It ensures that on an average more than $50\%$ of the expected blocks (one half of $\delta$) are added to the branch in which the block is present for each tournament (i.e., a majority of the network is contributing to the current branch continuously for a set of previous tournaments) and a hidden chain heavier than the current chain does not exist. It also makes sure that a fork cannot reverse a confirmed transaction because it needs more than $\delta/2$ number of blocks per tournament for almost $2x$ number of tournaments. In both cases, greater than $50\%$ of the nodes (i.e., majority of the network) are required to extend the malicious chain. Therefore, as long as the majority of users in Colosseum are honest, a valid transaction cannot be reversed.

    \subsection{Byzantine faults}
        Nodes in the system can go rogue. They can perform malicious activities which favors them to take control of the system. CDAG depends on Colosseum for the selection of block proposers and can reasonably affect the progress and security of CDAG. Users play three roles in Colosseum: Player, Validator and Keeper. As a player, a malicious node can try to cheat and win, whereas validators and keepers can avoid their tasks of validation and verification to make other nodes suffer.\\
        
        \noindent\textbf{Malicious Player.} \textit{Players} in Colosseum need to find other players and contest in a match. These players can try to play multiple matches for the same round with different opponents in order to increase their chances of moving to the next round. Keepers in Colosseum can easily detect such matches and can alert other nodes in the system with verifiable proof. Result of the matches of the same round will land up with the same keeper replicas nodes and results in verifiable proof propagated in the network. \\

        \noindent\textbf{Malicious Validator.} A \textit{validator} in Colosseum collects the proposals from the players, compute the result and acknowledge with a Proof-of-Win. He is also responsible for sending the result to the \textit{keeper}. In the case when an adversarial node is selected as the validator he may choose, 1) Not to reply, 2) Send result to players but not the keepers, 3) Send result to the keepers but not the players, 4) Send result to only one of the players and not to keepers, 5) Send delayed reply to the players.
        
        In the first case both the nodes will timeout, check with the keepers (both their and opponent's) if they got the result and move forward to play the same round with new competitors on being unsuccessful. The second case can result in keepers voting against the players for not receiving the proof-of-win. To prevent it nodes on receiving the result forward it to some of their keepers and the keepers gossip it forward. The third case is indistinguishable from the first case for the players as they will timeout and poll the keepers for the result, but this time the query will be successful. Fourth further gets divided into two subcases based on the player receiving the result is a winner or a loser. If the player is a winner, he will forward the certificate to its keepers such that they do not vote against him. The second player will find this Proof-of-Win (because same result certificate for both the players) from the keepers and will not proceed forward to play the same round again. On the other hand, if the receiving node was a loser and does not share the result with the keeper, both the nodes can play the same round again as nobody in the system knows about the match other than the losing node and the malicious validator. It is not fair to both the players but does affect the safety of the protocol. Also, it is similar to the first case when none of the players receives the result and both of them proceed. Also, random validator selection makes sure that such scenarios do not occur frequently.
        
        The fifth case is more hazardous than others for honest players in Colosseum. A player waiting for a response from the validator may timeout and start finding a new opponent for the same round again. This situation is indistinguishable from the case when a player is trying to play multiple matches for the same round and can lead to the detection of an honest node as malicious. To prevent this situation of having a false-positive an honest node proceeds if he wins both the matches, i.e., he is the winner of the match whose result is received after the timeout and also wins the second match played. In case he lost the first match, he was intended to stop playing, which is fair and happens the same in this case. However, if he wins the first match and loses the second, he still needs to stop moving forward to not to get marked as malicious by the participants. Whereas, a node with both wins can play further and is not considered malicious. It is because, if a player knows that he has won, he will not try to play for the same round again as doing this can mark it as malicious if he loses by any chance. Therefore, keepers consider it as an honest mistake and do not send alerts in the network. However, it does not protects the malicious players trying to play multiple matches for a round as they need to win all of them to proceed in the tournament. \\

        \noindent\textbf{Malicious Keeper.} \textit{Keeper} plays the role of storing the state of nodes on the network, verifying their previous state and notify the network of any unauthorized activity. Multiple random nodes are selected as the keepers of a match to have a fault tolerant storage. However, one or more of these keepers may get compromised and do not perform their tasks correctly. A keeper may, 1) Not verify a match whose PoWin is received, 2) not store the state with it, 3) send false alerts about unauthorized plays, 4) vote against matches of honest nodes. Most of these problems for the keepers of Colosseum gets solved because of their replication across the network. The honest counterparts will verify matches not verified by a malicious keeper. The state is stored at multiple nodes distributed over the entire network, and therefore, votes of a few dishonest nodes do not affect the final result for the players. Also, the keepers cannot generate false alerts as an alert requires proof of unauthorized play (multiple Proof-of-Win for the same round or Proof-of-Win for a higher round number of a player already lost in previous rounds).\\
        
        Moreover, malicious nodes can masquerade as keepers and validator nodes for the matches because all the nodes in Colosseum are arranged in a DHT and are generally responsible for storing data for a range of keys. They can try to generate PoWin's and cast votes on behalf of others. Colosseum uses a simpler form of DHT where nodes know all the other nodes present on the ring, making it challenging to masquerade others. Also, an honest node can easily detect and alert the system against such messages as the expected churn rate for the network is less and the participants are mostly known. Furthermore, researchers have extensively studied these topics in the past, and there are various solutions like \cite{castro2002secure,wang2006myrmic,wang2012octopus}, any one or a combination of them can be used with our protocol to improve and secure the messages over the DHT.

    \subsection{Bypassing Distributed time Barrier}
        Recent works (e.g., \cite{biondo2018guard,brasser2017software}) suggest that bypassing the trusted execution environments (TEE) and undermining the verification process of PoET is possible. Colosseum assures that as long as a majority of the nodes in the network are honest, the protocol continues to work synchronously. 
            
        A match between two players depends on random validators and keepers. These nodes do not process a message if the difference between the tournament number of the user's currents state and any received message is more than one. Therefore, it is difficult for malicious players to enter a tournament not started for others. Dependency on other nodes indirectly forces everyone to move collectively. Thus, the model supports the idea of the distributed barrier and prevents adversarial nodes from undermining its entirety.
            
    \subsection{Liveliness vs Safety}
        A consensus protocol cannot guarantee both liveliness and safety for asynchronous networks, as stated by the FLP impossibility result \cite{fischer1982impossibility}. CDAG chooses liveliness over safety. It guarantees liveliness as long as the majority of the selected proposers for the tournaments of Colosseum are honest and extend the heaviest chain of C-Blocks. A transaction in CDAG gets committed if  the majority of the network is working towards extending the heaviest chain, as described in section \ref{txnCnfrm}, and more than one branch cannot have majority of blocks simultaneously makes sure that eventually blocks get confirmed and nodes reach to a consensus, but not in the case of network partitions. However, safety is not guaranteed by CDAG because temporary forks can occur, i.e., two honest nodes can have different truths at some point in time but get resolved irreversibly.
            
    % \subsection{Progress}
    %     CDAG must progress, i.e., blocks of transactions should keep getting added to the ledger.  In protocols where a single leader or a small group of validators continue proposing blocks, it becomes easy for attackers to locate the targets and execute a DDoS~\cite{ddos} attack. Therefore, CDAG progresses by having multiple block proposers for each tournament. A new set of proposers qualifies for each tournament which makes it hard for an attacker always to be able to attack all the proposers before they propose a block.
\section{Implementation}
    \begin{table}
        
    \centering
        \resizebox{\columnwidth}{10ex}{%
         \begin{tabular}{||c |l | c||} 
             \hline
             Parameter & Meaning & Value \\ [0.5ex] 
             \hline\hline
             $\alpha$ & number of wins required to propose a block & 4\\
             \hline
             $K$ & keeper replication factor & 16 \\
             \hline
            %  $m$ & percentage of malicious nodes & 0-20 $\%$\\
            %  \hline
            %  $S$ & block size in MB & 1- 4\\
            %  \hline
             $\tau$ & Distributed barrier delay in sec & 20 sec\\
             \hline
             $B$ & number of buckets of transactions & 40\\
             \hline
             $F$ & minimum full confirmations to confirm a block & 3 \\ [0.5ex] 
             \hline
        \end{tabular}%
        }
         \vspace*{0.1in}
        \caption{Tunable parameters in the protocol}
        \label{table:one}
        \vspace*{-0.1in}
    \end{table}
    
    % \anonymize{}
    % ~\cite{reddy2006vishwa}
        We implemented  CDAG and Colosseum in JAVA consisting of all the elements that are significant to analyze its performance. For the underlying network, we use Vishwa~\cite{reddy2006vishwa} a peer-to-peer grid computing middleware . It uses ZonalServer (trusted oracle in case of Colosseum) to manage the underlying structured network in Colosseum which forms a DHT and provides bootstrap information to newly joined nodes. Colosseum uses a similar network model and is, therefore, a perfect match.
        
        In our implementation user on receiving a message gossips it to three nodes (two random nodes from the routing table and an immediate successor) in the structured layer. All the messages are digitally signed and also contain the public-key certificates of the user (we can also broadcast public-key certificates initially to save the overhead). We use waiting to simulate the behavior of PoET for the distributed time barrier. A block contains the header, PoWin and dummy data instead of actual transactions.
        
        Table~\ref{table:one} shows all the tunable parameters in the protocol with their default values unless specifically mentioned for experiments. We set the minimum number of full-confirmations required to consider a block as confirmed to 3. The keeper replication factor is set to 16 for all the experiments but depends on the scale of the system.
    
     \begin{figure*}[!ht]
          \begin{subfigure}[b]{0.33\textwidth}
            \includegraphics[width=\linewidth]{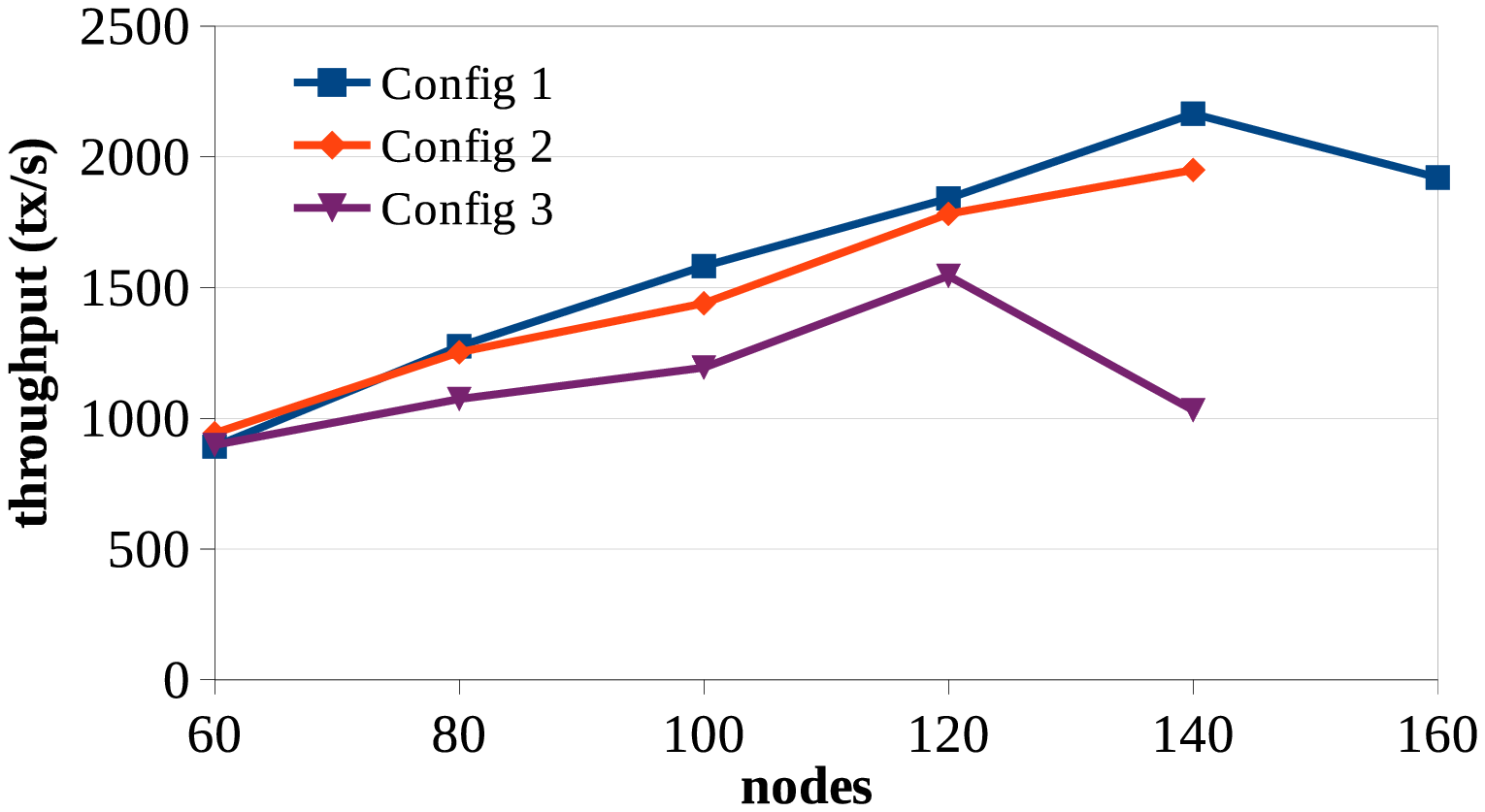}
            \caption{}
            \label{fig:a1}
          \end{subfigure}
          \begin{subfigure}[b]{0.33\textwidth}
            \includegraphics[width=\linewidth]{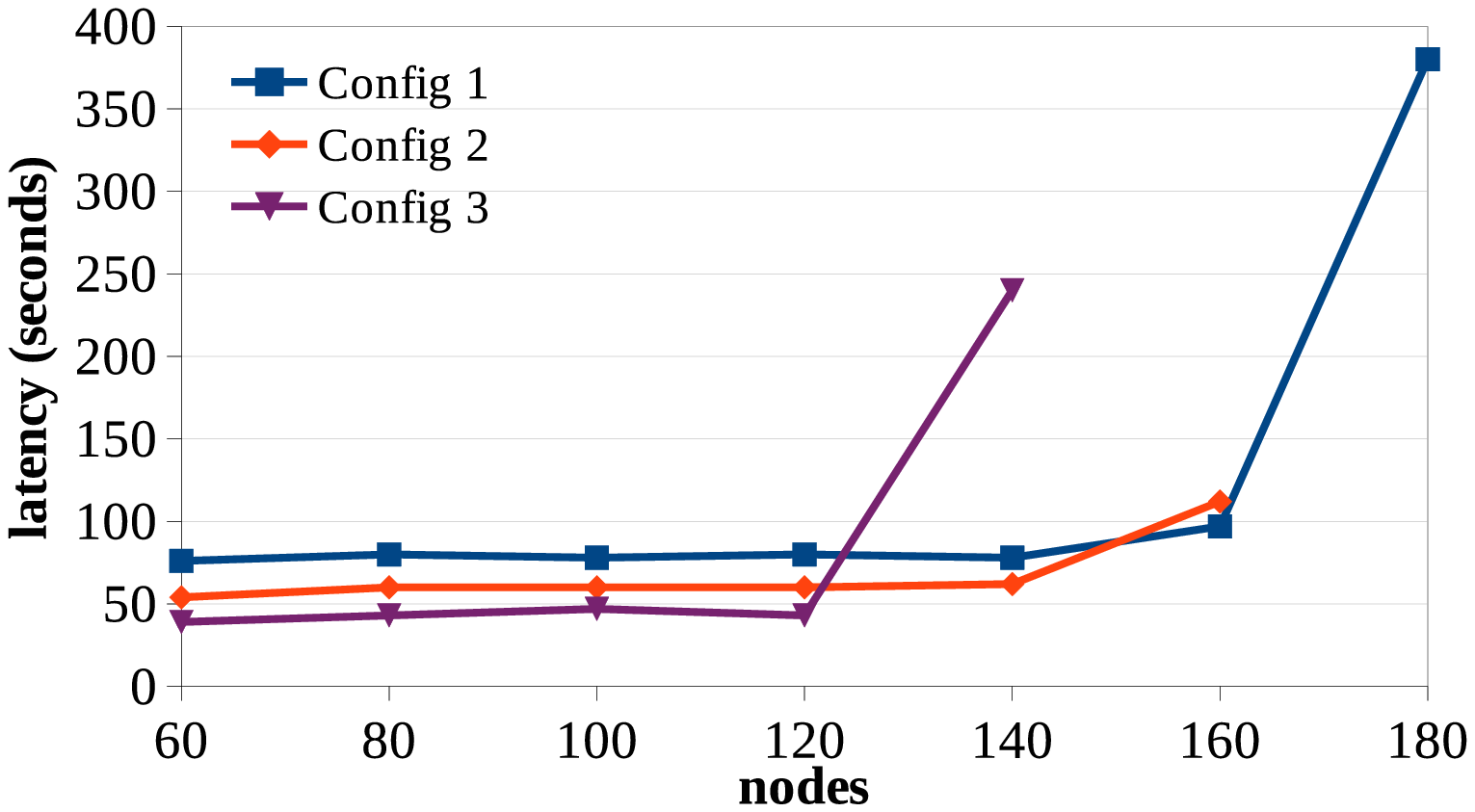}
            \caption{}
            \label{fig:a2}
          \end{subfigure}
        \begin{subfigure}[b]{0.33\textwidth}
            \includegraphics[width=\linewidth]{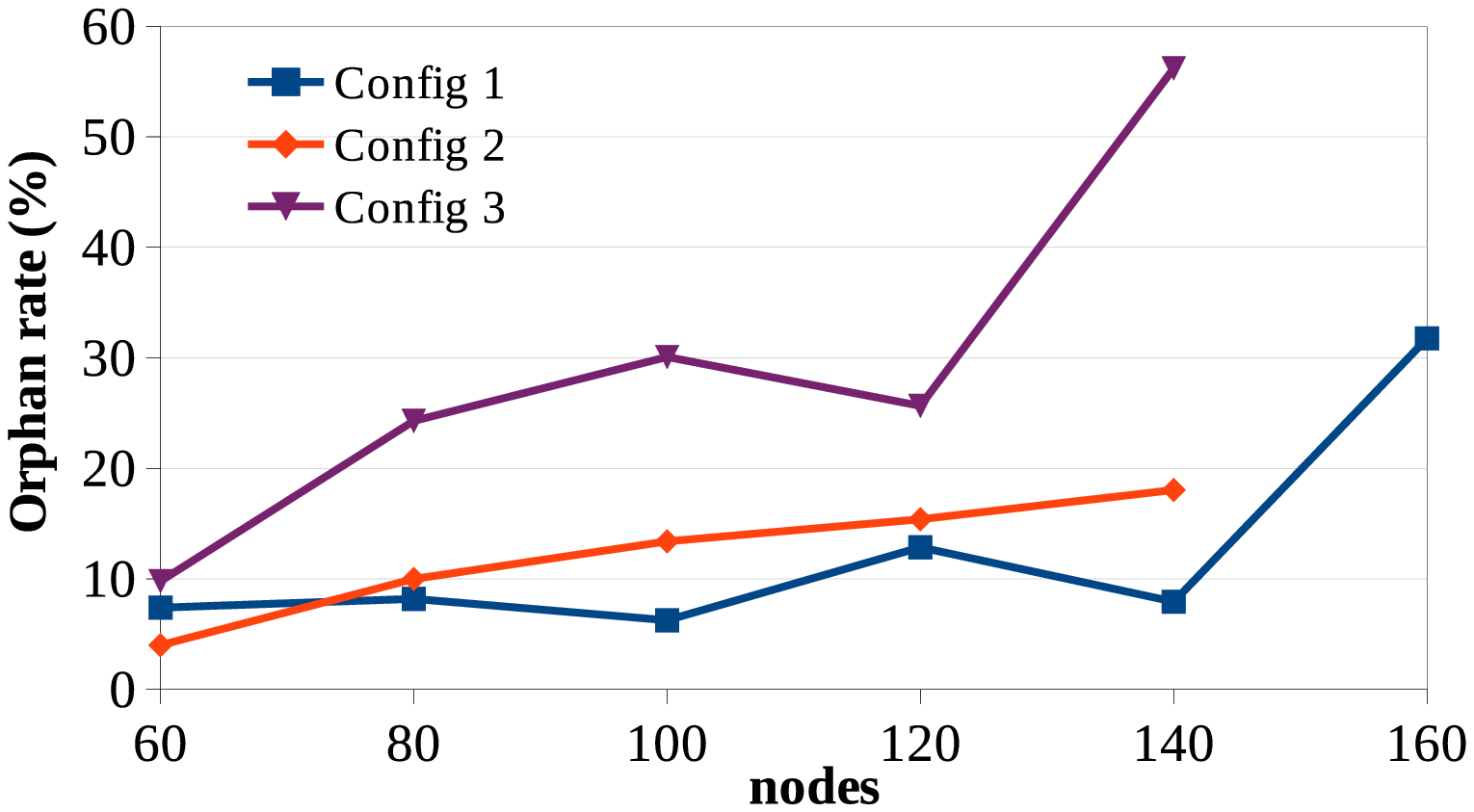}
            \caption{}
            \label{fig:a3}
          \end{subfigure}
        \caption{Throughput, Latency and Orphan rate of CDAG for $\alpha = 3$}
    \end{figure*}
        
    \begin{figure*}[!ht]
          \begin{subfigure}[b]{0.33\textwidth}
            \includegraphics[width=\linewidth]{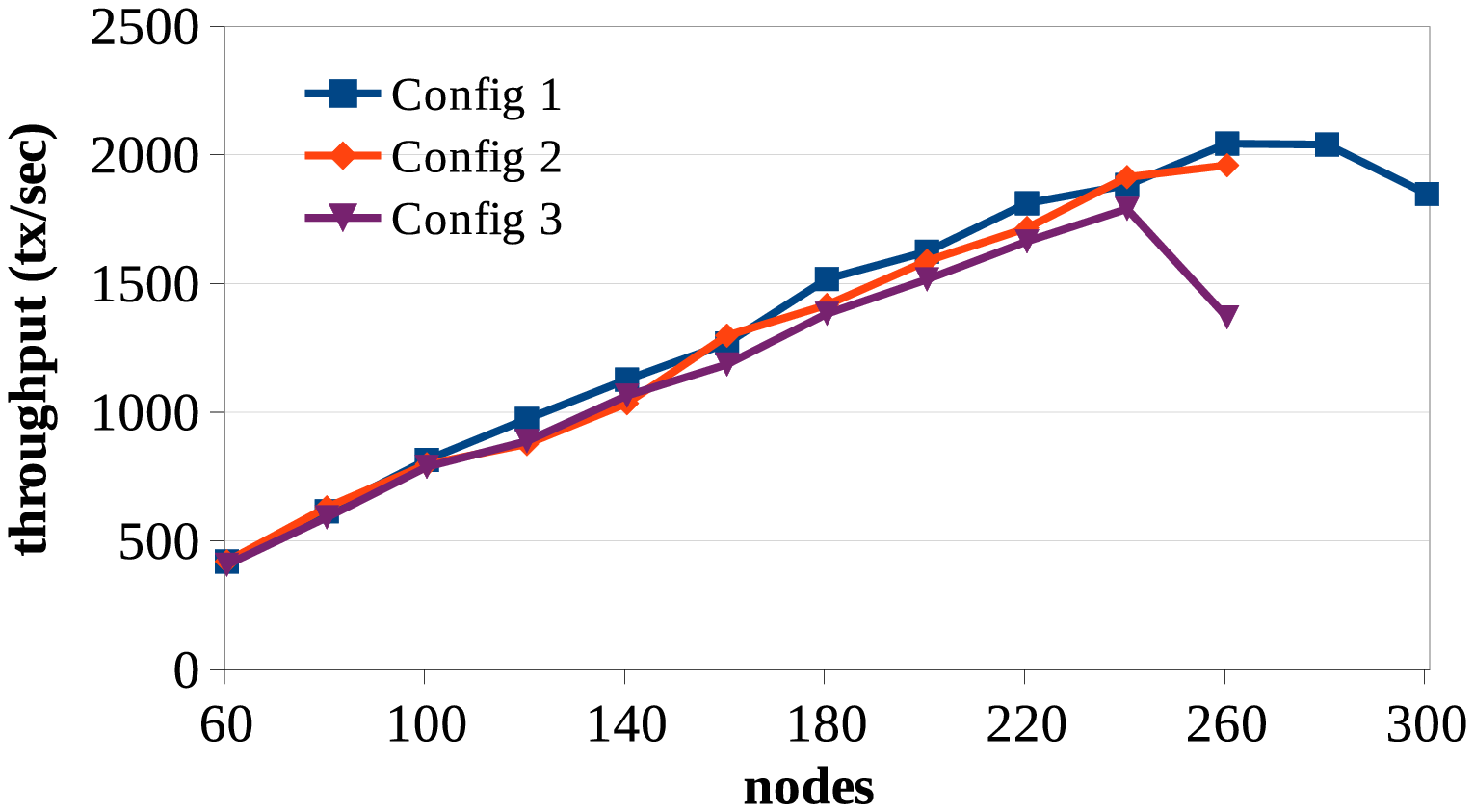}
            \caption{}
            \label{fig:b1}
          \end{subfigure}
          \begin{subfigure}[b]{0.33\textwidth}
            \includegraphics[width=\linewidth]{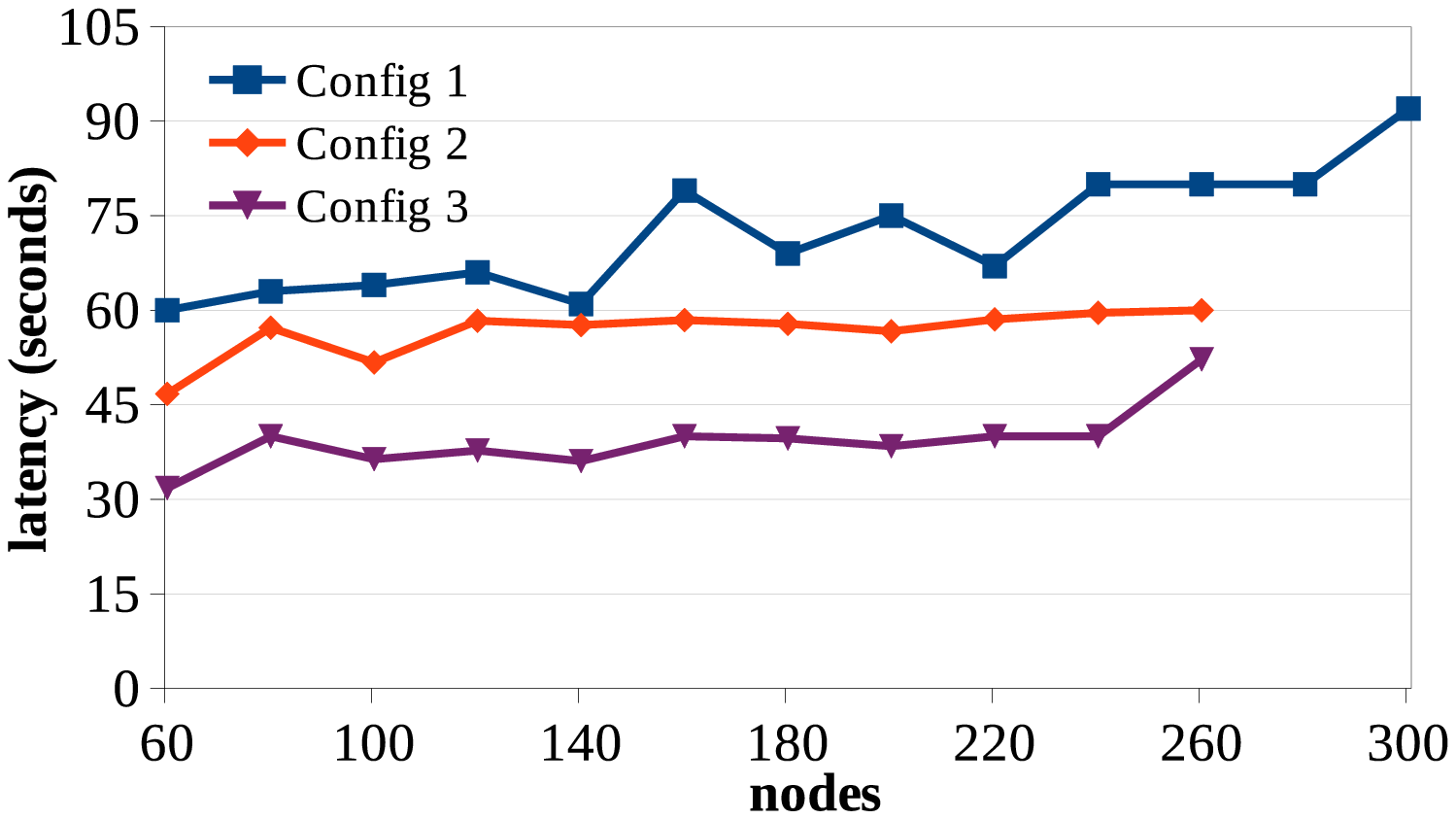}
            \caption{}
            \label{fig:b2}
          \end{subfigure}
        \begin{subfigure}[b]{0.33\textwidth}
            \includegraphics[width=\linewidth]{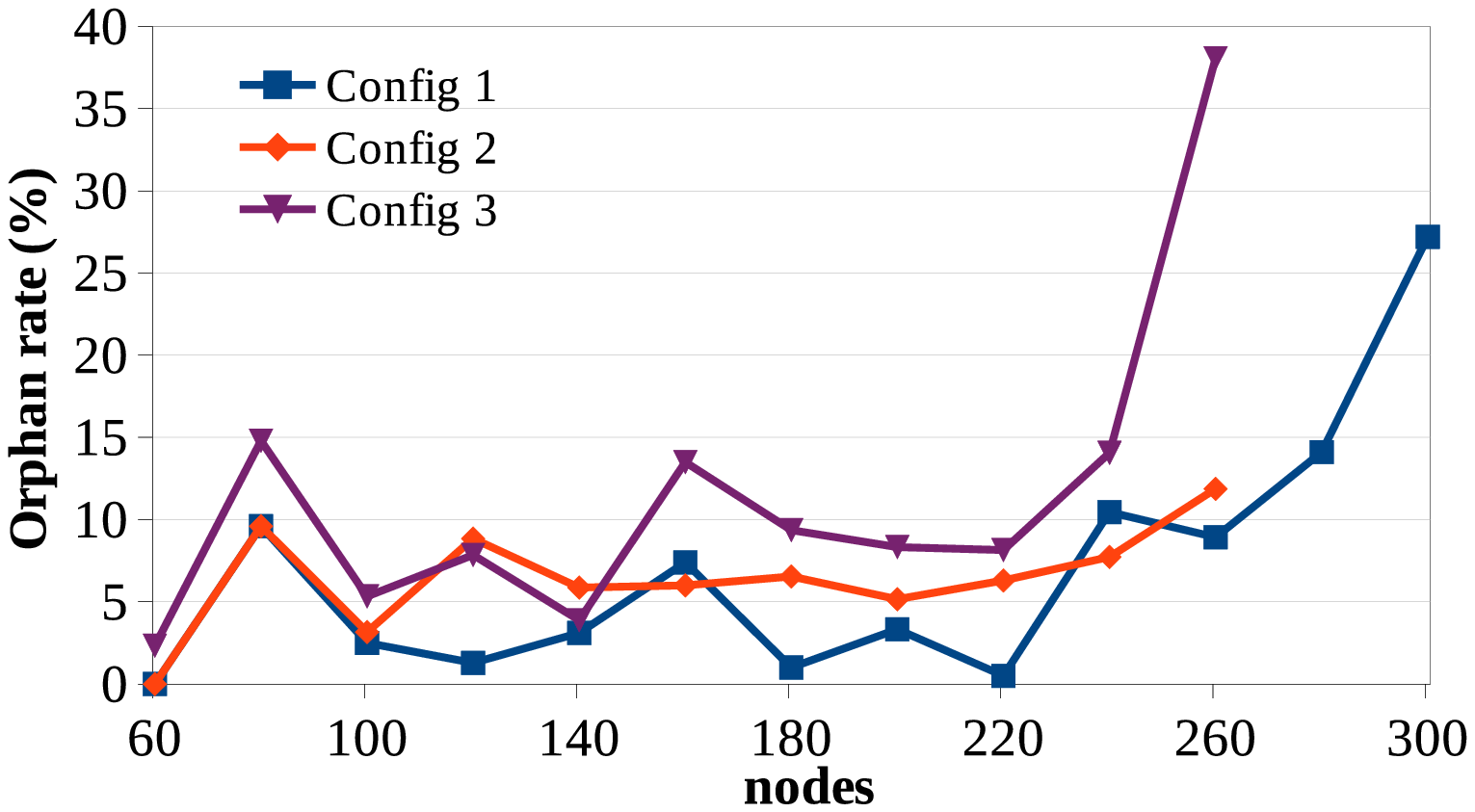}
            \caption{}
            \label{fig:b3}
          \end{subfigure}
          \caption{Throughput, Latency and Orphan rate of CDAG for $\alpha = 4$}
    \end{figure*}

\section{Evaluation}
        
        We evaluate the performance of CDAG for different sizes of network and vary the percentage of malicious users present in it. The experiments measure the throughput and show the variation in the latency and orphan rate of the protocol for three different configurations, as shown in table \ref{table:two}. The results demonstrate that how CDAG scales with the number of nodes by tuning some hyper-parameters and still give stable results.  We also measure the performance of CDAG under the influence of adversaries in the network and show its effect on some of the performance parameters.
        
        \begin{table}[h]
            \centering
            \begin{tabular}{||c |c | c||}
                \hline
                 Configuration & Block Size & Time slot ($\tau$) \\ [0.5ex] 
                 \hline\hline
                Config 1 & 1 MB & 20s \\
                \hline
                Config 2 & 0.75 MB & 15s \\
                \hline
                Config 3 & 0.50 MB & 10s \\[0.5ex]
                \hline
            \end{tabular}
            
            \vspace*{0.1in}
            \caption{Different configurations used in experiments}
             \vspace*{-0.1in}
            \label{table:two}
        \end{table}
        
        We deploy CDAG on Google Cloud Platform \cite{gcp} using Google Kubernetes Engine~\cite{gke}. The Kubernetes cluster consists of multiple n1-highmem-16 instances, each with 16vCPUs and 104 GB memory. Participants of CDAG are encapsulated in docker containers~\cite{merkel2014docker} and deployed on the cluster with an upper and lower limit of twelve and eight docker containers per instance and bandwidth constraint of 25 Mbps per docker container. We also make sure that multiple dockers on the same instance also adhere to the bandwidth constraints. Users propose blocks of different sizes for different set of configurations with 1 MByte equivalent to 2500-3000 transactions. Each experiment runs for at least 30 time slots and the graphs in this section plots an average over all of them.

        % \begin{figure}
        %     \centering
        %     \includegraphics[width=25em,height=14em]{graphs/graph1_3.eps}
        %     % \vspace*{-0.1in}
        %     \caption{Throughput of Colosseum for different values of $\alpha$, with 60 to 300 users in the structured layer}
        %     \label{fig:throughput}
        % \end{figure}
        
        \subsection{Throughput}
            To measure the throughput (transactions/sec) of the protocol, we vary the number of nodes in the network for multiple values of $\alpha$ and all the three configurations as mentioned in table \ref{table:two}. Both the graphs \ref{fig:a1} and \ref{fig:b1} shows that for a fixed value of $\alpha$ the throughput of the system increases initially with the number of nodes and then starts decreasing. The initial increase is because the number of blocks for a tournament of Colosseum $\delta$ is directly proportional to the number of nodes, as shown in equation~\eqref{eq:five}. More number of nodes results in a higher degree of parallelism for CDAG and thus, increases the throughput. After a certain point, the throughput starts decreasing for all the different configurations in \ref{fig:a1}, \ref{fig:b1} due to the generation of a large number of simultaneous blocks. Bandwidth constraints and network congestion do not provide them enough time to reach everyone in the network. Also, it is evident in both the graphs that the configuration with smaller time slot $\tau$ (e.g., Config 3) saturates before than the ones with a higher time slot (e.g., Config 1) even though the size of the block also increments in the same ratio. 
            
            Across \ref{fig:a1} and \ref{fig:b1} the throughput of CDAG decrease for the same network configuration. It is because, with the increase in the value of $\alpha$, the number of qualifying proposers $\delta$ for a time slot decreases, as inferred from equation~\eqref{eq:five}. However, the saturation point shifts to a higher number of nodes for larger values of $\alpha$. It allows CDAG to scale with the number of nodes. The value of $\alpha$ can be configured to achieve the maximum throughput for a given configuration of the network and can change with time similar to setting the hardness of the cryptographic puzzle in Bitcoin.
            
            Results show that CDAG achieves a throughput of approx 2100 transactions/sec  with a network size of approx 300 nodes and can also scale further. It is 28.57\% greater than the on-chain transaction throughput of 1500 transactions/sec \cite{ripple} of ripple with the validator set ranging from 30 to 50 peers \cite{rippleNet}. On the other hand, Hyperledger~\cite{androulaki2018hyperledger} shows a throughput of 3500 transactions/second for about 100 peers. However, it uses very highly configured peers with 16vCPUs and 1Gbps network link in the experimental setup, unlike CDAG which is tested using very nominal resources as discussed earlier in this section.
            
        \subsection{Latency}
            Figure \ref{fig:a2}, \ref{fig:b2} shows the block confirmation time for different configurations (ref table~\ref{table:two}) with the increase in the number of nodes in the network. It is evident that the block confirmation time for configurations with smaller $\tau$ is generally lower in comparison to those having a large value of $\tau$. It is because the minimum confirmation time for a block is \textit{ F * $\tau$}, where F is the minimum \textit{full-confirmations} required to confirm a block and $\tau$ is the minimum time to get one full-confirmation.  
            
            Network saturation also affects the latency of the protocol, similar to what was observed for throughput. Increasing the number of block proposers (results from increasing the number of nodes for a particular $\alpha$) beyond a limit lowers the percentage of blocks getting added to the ledger for each slot, resulting in an increase in the number of slots required for a block to achieve one full-confirmation and therefore, increasing the latency. Between \ref{fig:a2} and \ref{fig:b2}, the network with a larger $\alpha$ has much more stable latency values for larger network sizes as a result of the lower network traffic due to lesser blocks in each slot. The saturation point also moves forward for a higher value of $\alpha$. It indicates towards a scalable blockchain protocol with acceptable latency values in the order one minute. 
            % Also, the latency does not increase in the presence of conflicting transactions in comparison to other blockDAG protocols, providing consistency to the ledger.
            
            % However, before the saturation point the latency for the configuration with smaller time slot is much lower without compromising on the throughput. There is a trade-off between stability and latency, and thus, the hyper-parameters can be tuned to achieve optimal results.
            
            \begin{figure*}[ht]
            \begin{subfigure}[b]{0.32\textwidth}
            \includegraphics[width=\linewidth]{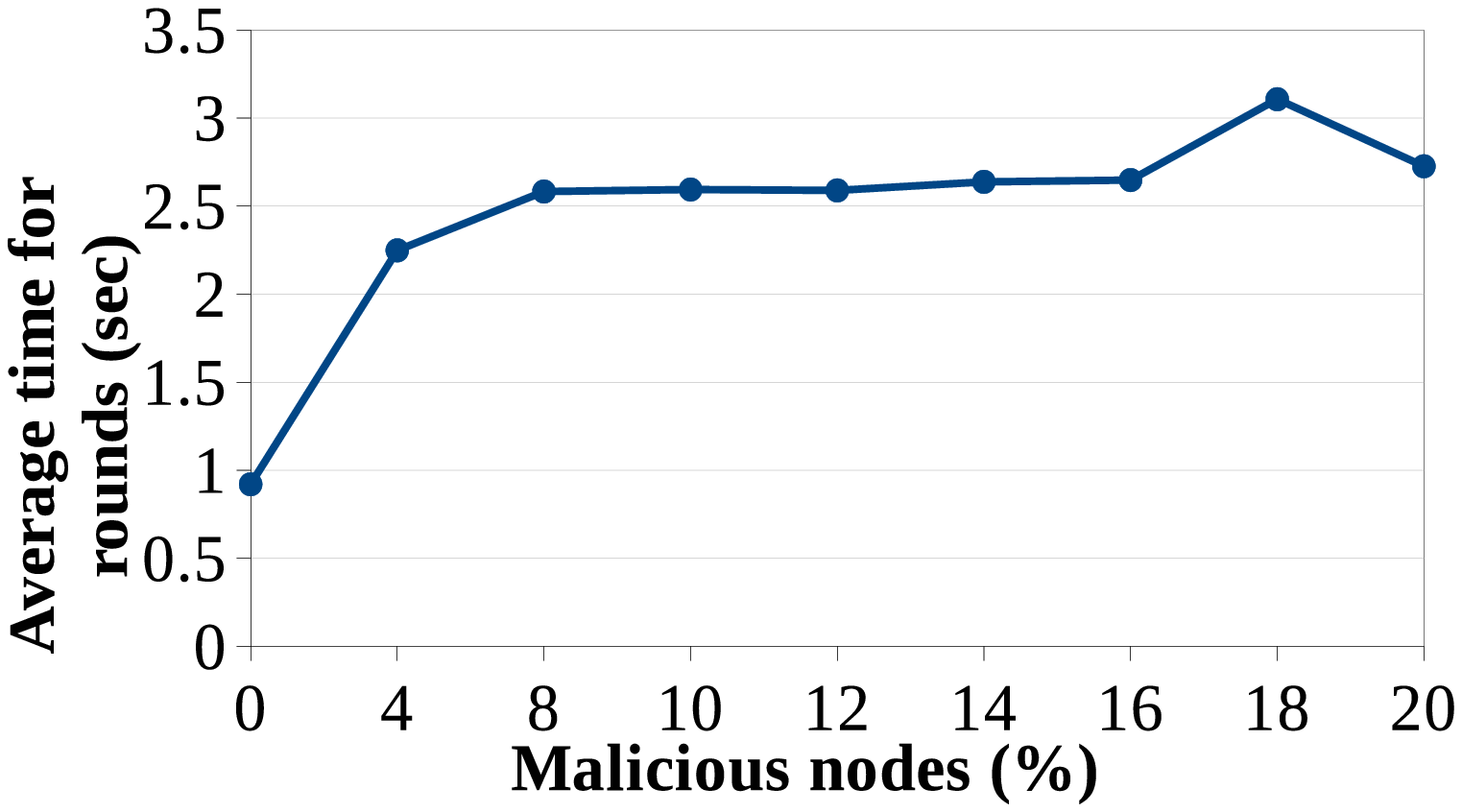}
            \caption{}
            % \caption{Average number of blocks in each C-Block with a varying fraction of malicious users, out of a total of 200}
            \label{fig:time}
          \end{subfigure}
          \begin{subfigure}[b]{0.32\textwidth}
            \includegraphics[width=\linewidth]{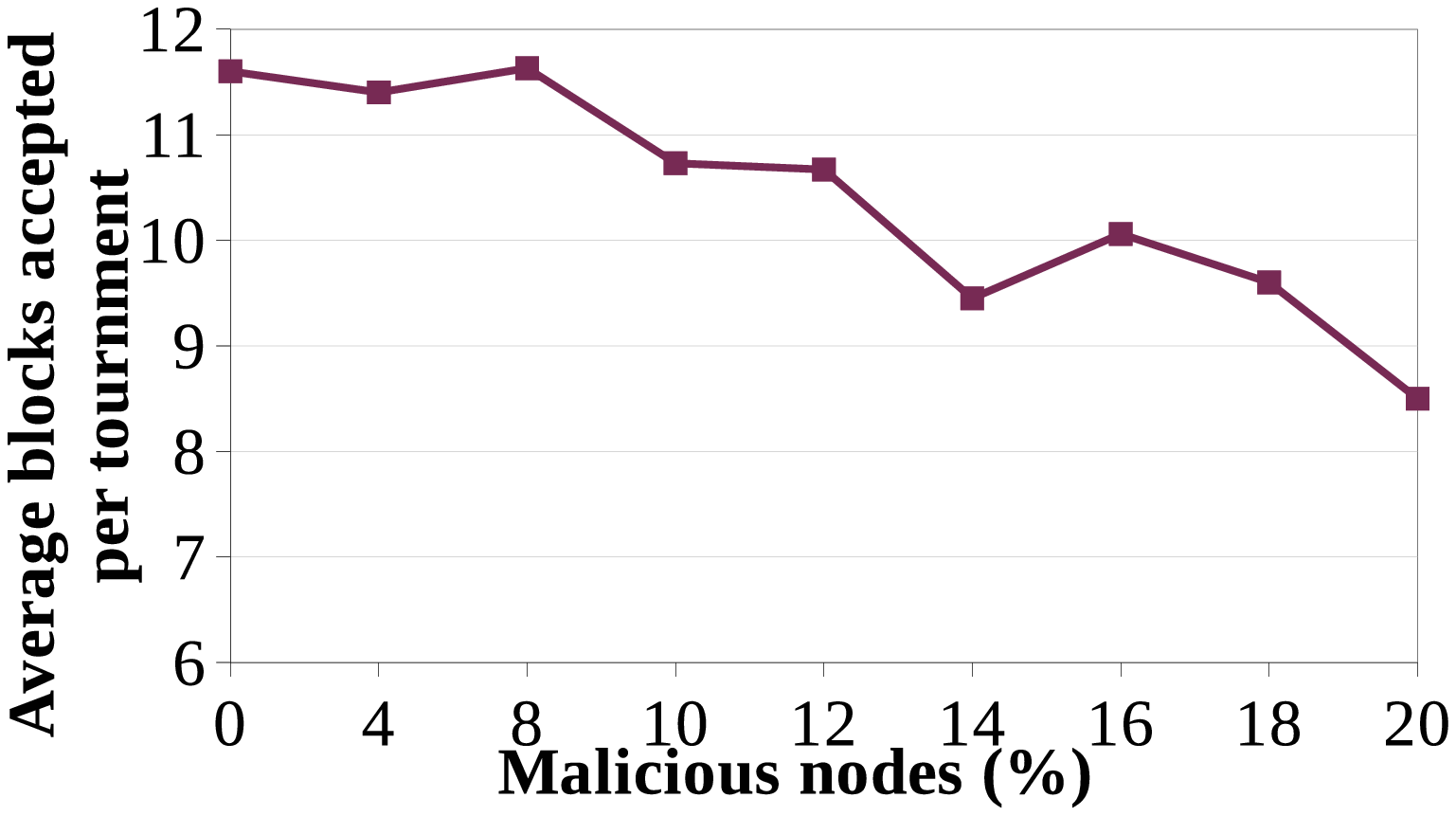}
            \caption{}
            % \caption{Average time for one round of the tournament with a varying fraction of malicious users, out of a total of 200}
            \label{fig:blocks}
          \end{subfigure}
          \begin{subfigure}[b]{0.32\textwidth}
            \includegraphics[width=\linewidth]{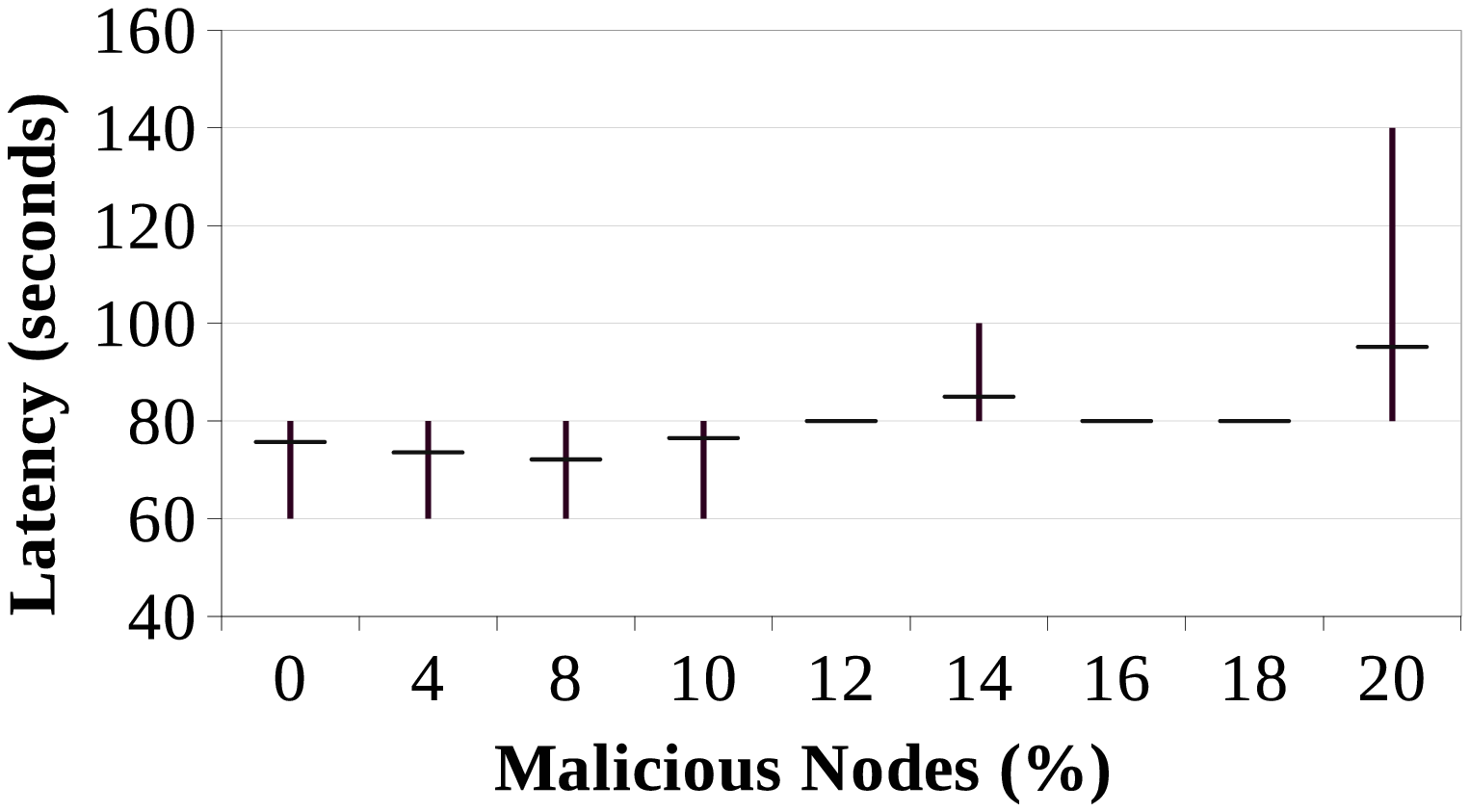}
            \caption{}
            % \caption{Min, Avg and Max transaction latency with a varying fraction of malicious users, out of a total of 200}
            \label{fig:latency}
          \end{subfigure}
          \caption{For a varying fraction of malicious users, (a) Average time for one round of the tournament, (b) Average number of blocks in each C-Block, (c) Min, Avg and Max transaction latency }
        \end{figure*}

        \subsection{Orphan rate}
            
            Orphan rate of the protocol shows a behavior similar to the latency of CDAG since both of them are correlated. Latency in CDAG is derived from the percentage of blocks getting added to the main chain with respect to the maximum, which defines orphan rate. In figure~\ref{fig:a3} and \ref{fig:b3} the orphan rate is initially in a stable range with the increase in the number of nodes and then shoots up because of the network congestion similar to figure~\ref{fig:a2} and \ref{fig:b2}.
            
            The duration of the time slot $\tau$ significantly affects the orphan rate of CDAG as providing sufficient time for the blocks to scatter in the network better utilize the network resources. Figure~\ref{fig:a3} and \ref{fig:b3} shows that the orphan rate is largely lesser for the configuration with the higher values of $\tau$ because of the higher stability achieved by the network for each slot. A larger $\tau$ allows more number of concurrent blocks to reach everyone in the network and get added to the heaviest chain, resulting in a drop in the orphan rate. Some inconsistencies in the initial part of graphs are because of a few blocks getting generated for each slot and the randomness involved in the bucket selection process while the generation of blocks. Selection of similar buckets by the proposers in such cases can have a significant effect on the percentage of discarded blocks in comparison to stages when the block frequency is high.

        \subsection{Misbehaving Users}
            We show that Colosseum can handle malicious users by randomly adding some percent of them in the network. These users defy the protocol by not performing any of the tasks of validators and keepers, but themselves play to propose blocks. Figure~\ref{fig:time}, \ref{fig:blocks} show the variation in the average time for the rounds of Colosseum and the number of blocks getting added to CDAG with the increase in the number of malicious players. The average time for individual rounds of a tournament increases with the number of malicious users and the average number of blocks accepted per tournament decrease. It is because the number of matches dropped by malicious validators increase and more blocks get discarded due to rise in the false-negative votes against matches of honest players by malicious keepers. 
            A little inconsistency in results is because of the randomness involved in placing the nodes over DHT and selecting validators and keepers for the matches.
            
            Figure~\ref{fig:latency} shows the minimum, average and maximum latency of Colosseum and its variation with the number of malicious users. It is evident that Colosseum confirms transactions in less than two minutes (120 seconds) when approx. 20\% of validators and keepers are not performing their tasks honestly.

\section{Limitations and Future Work}
        This paper discussed a new structure to store transactions over a distributed ledger similar to the blockchain. The structure suffers from temporary forks even if users propose non-conflicting blocks. It can be due to smaller time slots for the barrier or low network bandwidth. A blockDAG kind of approach on top of CDAG to converge multiple non-conflicting forked chains can help to decrease the orphan rate further. 
        
        We do not discuss optimal configuration setting for CDAG to achieve maximum throughput and minimum latency. A detailed analysis of the configuration parameters and discussion on the trade-offs can be helpful to gain performance and better utilize the resources. One can also explore game theory techniques to have easily verifiable multi-player distributed games to filter out the nodes and select the block proposals more efficiently and independently.
        
\section{Conclusion}
    CDAG is a new distributed data-structure to store transactions as an alternate to blockchain and blockDAG. It scales well with the number of nodes while maintaining the consistency in its performance. It uses Colosseum to select multiple block proposers in each time slot and use different buckets of transactions to generate non-conflicting blocks. Unlike other DAG based protocols, CDAG provides immediate finality to the blocks of the ledger and has a totally ordered structured by design. It is designed to have lower orphan rates for real-time scenarios and support smart contracts. Experimental results for Colosseum demonstrate that it achieves a throughput of more than 2000 tps and confirm transactions in less than two minutes.
    
    % Colosseum is a new permissioned blockchain protocol, both fair and fast. It scales well with the number of nodes while maintaining the consistency in its performance. Colosseum introduces CDAG to record transactions over the distributed ledger with the capability to commit multiple blocks simultaneously. The distributed barrier across the network helps to achieve step synchronous tournaments even in the presence of malicious users and is the key to control the orphan rate. Experimental results for Colosseum demonstrate that it achieves throughput of more than 2000 tps and confirm transactions in less than two minutes.

% The next two lines define the bibliography style to be used, and the bibliography file.
\bibliographystyle{ACM-Reference-Format}
\bibliography{main.bbl}

\end{document}